\documentstyle[preprint,prd,aps]{revtex}
\begin{document}
\draft
\title{Pion interferometry with pion--source-medium interactions}
\author{M.-C. Chu$^1$, S. Gardner$^2$, T. Matsui$^{2,3}$, and R.
Seki$^{1,4}$}
\address{
$^1$ W.~K.~Kellogg Radiation Laboratory, California Institute of
Technology\\
Pasadena, California 91125\\
$^2$ Nuclear Theory Center, Indiana University, Bloomington, Indiana
47408\\
$^3$ Yukawa Institute for Theoretical Physics, Kyoto University,
Kyoto 606, Japan\\
$^4$ Department of Physics and Astronomy, California State
University\\
Northridge, California 91330\\}
\date{\today}
\maketitle

\begin{abstract}
An extended pion source, which can be temporarily created by a high
energy nuclear collision, will also absorb and distort the outgoing
pions. We discuss how this effect alters the interferometric pattern
of the two-pion momentum correlation function. In particular, we show
that the two-pion correlation function decreases rapidly when the
opening angle between the pions increases. The opening-angle
dependence should serve as a new means of obtaining information about
the pion source in the analysis of experimental data.
\end{abstract}

\tightenlines
\narrowtext
\section{Introduction}
\label{secintro}

Hanbury-Brown-Twiss interferometry of identical particles emitted
from a chaotic source has long been used to infer the spatial and
temporal characteristics of that source
\cite{hbt,gglp,kp,shur,yk,gkw,pratt,bau}. Although originally
conceived in astronomy to the end of extracting stellar radii
\cite{hbt}, the technique has found extensive application as a
diagnostic tool in heavy-ion collisions \cite{yk,gkw,pratt,bau,bert}.
In most practical applications, the outgoing pions are assumed to
suffer only weak final-state interactions. For instance, the
interaction between the two detected pions is usually neglected, and
a simple plane wave solution is used to calculate the two pion
momentum correlation function, although in the case of charged pions,
it is multiplied by the Gamow factor to correct for their long-range
Coulomb force.

In this paper, we discuss another type of final-state interaction,
wherein each pion suffers in its exit from the source region. Namely,
{\it the interaction of the pions with the ``source'' itself}. Since
a pion source also absorbs and distorts pions, a source with finite
extent in space and time will alter the outgoing pion waves in the
source region. In particular, if the pion source is a droplet of
``quark-gluon plasma,'' it may be regarded as a black-body of pions
--- and of all other hadrons --- so that it casts its shadow on the
paths of the outgoing pions emitted from the plasma surface. In a
classical picture, this effect may be taken into account by
introducing a strong correlation between the pion momentum and its
emission points, so that the pions are emitted only in outward
directions, an effect similar to that caused by collective flow
\cite{pgg,heinz}. Pions may also interact in a dense pion medium via
$\rho$ or $\omega$ resonance formation, for example, and change the
effective source size \cite{gp}. One can also include baryon
resonances explicitly in the pion source \cite{bdh}. Yet there is a
fundamentally more important effect which arises due to the quantum
mechanical nature of interferometry: {\it the distortion of the
interference pattern generated by the pion--source-medium
interaction}.

Here we study the pion correlation function in the presence of
pion--source-medium interactions, as represented by a local optical
potential, within the eikonal approximation. We calculate the
distortion of the pion wave function explicitly, thereby allowing for
a direct investigation of the quantum mechanical interference
effects. We show that both the real and the imaginary parts of the
optical potential give rise to non-trivial modifications: the former
induces additional interference effects, whereas the latter leads to
a suppression of the correlation as the opening angle between the
pions increases. We demonstrate that the opening angle-averaged
correlation function acquires a non-Gaussian shape due to these
interactions, even if the source distribution is Gaussian.

\section{Basic principles}
\label{secbasic}

Before proceeding to include the final-state interactions, we first
review the basic principles of two-particle interferometry. Such a
description is easily found in the literature; we present it here to
clarify its usual assumptions, as well as the manner in which we
include pion--source-medium interactions.

In essence, two particle interferometry results from a constructive
--- or destructive --- interference of the two amplitudes for the
emission of two identical bosons --- or fermions --- from two
independent sources. Suppose that there are $N$ such independent
sources, localized at the space-time points $x_i$ $(i=1,...,N)$, and
that the emitted particles do not interact with each other after
their creation. Let $\varphi_{\bf k} (x_i)$ be the amplitude that a
particle with momentum ${\rm k}$ is emitted from a source at $x_i$.
Then the pair amplitude $\Psi_{{\bf k}_1 {\bf k}_2} (x_i, x_j)$ that
two particles are created by the source at $x_i$ and $x_j$ and are
detected later with momenta ${\bf k}_1$ and ${\bf k}_2$ is written in
a product {\it Ansatz} as
\widetext
\begin{equation}
\Psi_{{\bf k}_1 {\bf k}_2} (x_i, x_j) = {1 \over \sqrt{2}}
e^{i(\delta_i+ \delta_j)}\left[ \varphi_{{\bf k}_1} (x_i)
\varphi_{{\bf k}_2} (x_j)
\pm \varphi_{{\bf k}_1} (x_j) \varphi_{{\bf k}_2} (x_i)
\right]\;.
\label{amp}
\end{equation}
\narrowtext
We have taken out the overall phase factors associated with the
arbitrary phase of each single particle amplitude, and since the
detected particles are on shell, we can label the pair amplitude with
just the three-momenta ${\bf k}_1, {\bf k}_2$. The first term is the
amplitude that the particle of momentum ${\bf k}_1$ is produced at
$x_i$ with the particle of momentum ${\bf k}_2$ produced at $x_j$.
The second term is needed to ensure the proper symmetry of the total
amplitude with respect to exchange of the particle coordinates ---
$(+)$ for boson pairs and $(-)$ for fermion pairs. The probability of
observing a particle with momentum $\bf k$ is given by
\begin{equation}
P_1 ({\bf k}) = {1 \over N} \left\langle \left\vert
\sum_i e^{i\delta_i} \varphi_{\bf k} (x_i) \right\vert^2
\right\rangle\;,
\label{p1}
\end{equation}
whereas the joint probability of finding two particles with momenta
${\bf k}_1$ and ${\bf k}_2$ is
\begin{equation}
P_2 ({\bf k}_1, {\bf k}_2) = {1 \over N(N-1)} \left\langle \left\vert
\sum_{i \ne j} \Psi_{{\bf k}_1 {\bf k}_2} (x_i, x_j) \right\vert^2
\right\rangle\;.
\label{p2}
\end{equation}
Note that $\langle \ldots \rangle$ indicates an average over the
source distribution. The two-particle momentum correlation function
is defined by
\begin{equation}
C_2 ({\bf k}_1, {\bf k}_2) \equiv
{P_2 ({\bf k}_1, {\bf k}_2) \over P_1({\bf k}_1) P_1({\bf k}_2)}\;.
\label{c2}
\end{equation}

With a few more assumptions, to be summarized below, the momentum
correlation function is related directly to the Fourier transform of
the source distribution, $\tilde\rho_{\tilde q} \equiv \int d^4 x
\rho (x) e^{i{\tilde q}x}$. Namely,
\begin{equation}
C_2 ({\bf k}_1, {\bf k}_2) = 1 \pm \left[ \left\vert
\tilde\rho_{\tilde q} \right\vert^2 /
\left\vert \tilde\rho_{{\tilde q}=0} \right\vert^2 \right] \;,
\label{c2'}
\end{equation}
where ${\tilde q} \equiv (E_1 - E_2, {\bf k}_1 - {\bf k}_2)$ is the
relative four-momentum. Hence Eq.~(\ref{c2'}) predicts that
\begin{equation}
C_2 ({\bf k}_1, {\bf k}_2) \to 2 (0) \qquad \qquad
\hbox{as} \qquad {\tilde q} \to 0
\end{equation}
for boson (fermion) pairs, whereas $C_2 \to 1$ as ${\tilde q} \to
\infty$ in both cases.

There are three assumptions, besides the {\it Ansatz} in
Eq.~(\ref{amp}), involved in deriving the above from Eq.~(\ref{c2}).
\begin{enumerate}
\item
The relative phases of particles created at different locations are
incoherent --- the phase average is random.
\item
There are no correlations in the source distribution.
\item
The particles do not interact after their emission.
\end{enumerate}
The first assumption is the most important of all: it allows us to
write the probability distribution in terms of an {\it incoherent
sum} of the individual probabilities. That is,
\widetext
\begin{eqnarray}
P_1({\bf k}) & = & {1 \over N}
\langle \sum_i \left\vert \varphi_{{\bf k}_1} (x_i) \right\vert^2
\rangle
= \int d^4 x \rho(x) \left\vert \varphi_{{\bf k}_1} (x)
\right\vert^2\;, \\
\label{p1'}
P_2 ({\bf k}_1, {\bf k}_2) & = &
{1 \over N(N-1)} \langle \sum_{i,j}
\left\vert \Psi_{{\bf k}_1 {\bf k}_2} (x_i, x_j) \right\vert^2
\rangle
\nonumber \\
& = & \int d^4 x_1 d^4 x_2 \rho_2 (x_1, x_2)
\left\vert \Psi_{{\bf k}_1 {\bf k}_2} (x_1, x_2) \right\vert^2\;,
\label{p2'}
\end{eqnarray}
\narrowtext
where the averages over the $N$ point sources are replaced by
integrals over smooth distributions. All the interference terms
containing the relative phase factor $e^{i(\delta_i - \delta_j)}$
drop out when averaged over the random phases $\delta_i$. Note that
the two terms in the bracket of Eq.~(\ref{amp}) have a definite
relative phase, so that their interference survives the random phase
averaging. The second assumption allows us to replace the
two-particle source distribution $\rho_2 (x_1, x_2)$ by the product
of the single particle source distributions:
\begin{equation}
\rho_2 (x_1, x_2) = \rho (x_1) \rho (x_2)\;.
\label{rho2}
\end{equation}
This factorization does not hold in the presence of fine structure in
the source, for example, in the multidroplet model of the mixed phase
\cite{psv}. Lastly, by the third assumption, the single particle
amplitude $\varphi_{\bf k} (x)$ is given as a solution of the free
Klein-Gordon or Dirac equation, as appropriate:
\begin{equation}
\varphi_{\bf k} (x) = c_{\bf k} e^{ik \cdot x}\;.
\label{green}
\end{equation}
Note that the normalization factors $c_{\bf k}$ cancel out in the
correlation function of Eq.~(\ref{c2}).

Finally, these assumptions lead to the expression \cite{yk,gkw,pratt}
\widetext
\begin{equation}
C_2({\bf k}_1,{\bf k}_2)
= {\int {\rm d}^4x_1\;{\rm d}^4x_2\;
\rho(x_1)\rho(x_2) \vert\Psi_{{\bf k}_1 {\bf k}_2}(x_1,x_2)\vert^2
\over{
\int {\rm d}^4x_1\; \rho(x_1) \vert\varphi_{{\bf k}_1}(x_1)\vert^2
\int {\rm d}^4x_2\; \rho(x_2) \vert\varphi_{{\bf k}_2}(x_2)\vert^2
}} \;,
\label{Cstart}
\end{equation}
where
\begin{equation}
\Psi_{{\bf k}_1 {\bf k}_2}(x_1,x_2)
= {1\over \sqrt{2}} \left[
\varphi_{{\bf k}_1}(x_1)\varphi_{{\bf k}_2}(x_2) \pm \varphi_{{\bf
k}_2}(x_1)
\varphi_{{\bf k}_1}(x_2)
\right] \;.
\label{eq12}
\end{equation}
\narrowtext

\section{Inclusion of pion--source-medium interactions}
\label{secfsi}

We now turn our discussion to the modification of the correlation
function due to the inclusion of final-state interactions. In
principle, there are two types of final-state interactions to be
considered: the mutual interaction between the detected pions, and
the interaction of the pions with the rest of the system. The former
effect may be included by solving the Bethe-Salpeter equation or the
Schr\"odinger equation for the two-body wave function \cite{yk,gkw}.

Here we examine the latter effect, the medium modification
\cite{gkw,gyu}. We incorporate it by adding a self-energy term to the
Klein-Gordon equation:
\begin{equation}
\left(\Box + m^2 + 2m U (x) \right) \varphi_k (x) = 0\;.
\label{kg'}
\end{equation}
We have assumed that $U(x)$ is a {\it local} optical potential. The
heuristic derivation of Eq.~(\ref{Cstart}) in Sec.~II presumes that
the pions do not interact after their emission. However,
Eq.~(\ref{Cstart}) still holds when a local one-body optical
potential is included in the construction of $\varphi_k(x)$ as long
as $U(x)$ does not support bound states or create real particle pairs
\cite{gkw}.

Many of the complications associated with the heavy-ion reaction are
lumped in the potential $U(x)$. For example, the pion source function
$\rho (x)$ appears explicitly in Eq.~(\ref{Cstart}), but the
dynamical response of the source to the interactions of the outgoing
pion with the source medium does not. This response is contained in
$U(x)$ --- it is the effective single-channel optical potential. The
final sum over the unobserved hadronic states implicit in
Eq.~(\ref{Cstart}) can still be effected independently of the pion
amplitudes in this picture. Furthermore, the initial state of the
source function is also not identified, and for a given dynamical
model description of the heavy-ion reaction, all allowed initial
states must be averaged. We presume, for now, that this averaging can
also be absorbed in $U(x)$.

The complexity of the optical potential is compounded by the time
evolution of the source. For example, as the source expands, the
relative momentum of the outgoing pion and of the source element with
which it interacts changes in time and thereby alters the pion's
interaction with the source. To illustrate this, consider an
expanding source that consists merely of nucleons. The strength of
the pion-nucleon interaction depends strongly on the relative
momentum between the pion and the nucleon with which it interacts.
Moreover, pion absorption by the source will depend on how two
nucleons in the source are correlated in time. The complexity of
$U(x)$ can be reduced by constructing an explicit dynamical model of
the heavy ion reaction, but simple physical insight into the effects
of $U(x)$ would be lost in such a task. Thus, we make the problem as
simple as possible, in order to extract the essential physics that
emerge as a consequence of the pion--source-medium interaction. To
this end, we first choose to ignore the additional complications that
arise from the time dependence of the source and consider merely a
static pion interaction with the source medium. Note that
Eq.~(\ref{kg'}) may then be interpreted as describing the
time-reverse of the usual scattering process: the pion returns from
infinity with momentum $-{\bf k}$ and is inversely scattered by the
optical potential $U(x)$. For pions with energies of a few hundred
MeV, $\varphi_{\bf k} (x)$ is given approximately by
\begin{equation}
\varphi_k (x) = c_{\bf k} e^{ikx - {i m \over k}
\int_z^{\infty} dz' U(x, y, z')}\;,
\label{glauber}
\end{equation}
where $(x,y,z)$ labels the pion production point. Note that
$z\equiv\bf{\hat{k}}\cdot{\bf r}$ and $z'$ parametrizes the straight
line path along ${\bf k}$ from $z$ to infinity.
Equation~(\ref{glauber}) is the eikonal approximation and agrees well
with elastic pion-nucleus scattering data in the energy regime we
consider \cite{hes}, when a realistic momentum-dependent potential is
assumed. We will use the function $\varphi_{\bf k} (x)$ in
Eqs.~(\ref{Cstart}) and (\ref{eq12}) to include the final-state
pion--source-medium interaction in the pion momentum correlation
function.

Moreover, we choose a simple model for the pion--source-medium
interaction --- a complex constant multiplies the source density
$\rho$:
\begin{equation}
{m \over k}U(r) = (\sigma_R + i\sigma_I)\rho (r) \;.
\label{potent}
\end{equation}
In general, $\sigma_R$ and $\sigma_I$ may depend on the momentum of
the pion with respect to the medium and on the temperature of the
medium as well. We shall assume that $\sigma_R$ and $\sigma_I$ are
simply constant, but we will examine a wide range of parameter values
in order to probe the sensitivity of the pions' momentum correlation
to this new effect. We can then write the distortion factor $g ({\bf
k}, {\bf r})$ as
\begin{equation}
g ({\bf k}, {\bf r})
= e^{-i(\sigma_R + i\sigma_I)t({\bf k}, {\bf r})} \;,
\label{gfactor}
\end{equation}
where the thickness function $t({\bf k}, {\bf r})\equiv
\int_z^{\infty}{\rm d}z'\; \rho({\bf b},z')$ is determined by the
direction of the outgoing pion and its production point ${\bf r}$.
The impact parameter ${\bf b}$ is defined by ${\bf r} - {\bf z}$,
where ${\bf z} = {\bf \hat{k}}\cdot{\bf {\hat r}}\, {\bf \hat{k}}$.
The real part of the potential generates a phase shift, whereas the
imaginary part represents a loss of flux. As we shall see below, both
factors can significantly modify the correlation function.

With the distorted single particle amplitude $\varphi_k (x)=
e^{ik\cdot x} g({\bf k},{\bf r})$ as per Eq.~(\ref{gfactor}), the
two-pion correlation function, Eq.~(\ref{Cstart}), becomes
\begin{equation}
C_2 ({\bf k}_1, {\bf k}_2) = 1 + \vert f ({\bf k}_1, {\bf
k}_2)\vert^2 \;.
\label{cf}
\end{equation}
The second term, generated by the interference of the two distorted
waves, is given by the square modulus of the overlap integral
\begin{equation}
f ({\bf k}_1, {\bf k}_2) = {1 \over N} \int d^3 r \rho ({\bf r})
g ({\bf k}_1, {\bf r})
g ({\bf k}_2, {\bf r})
e^{i{\bf q} \cdot {\bf r}} \;,
\label{f}
\end{equation}
where $N = \int d^3 r \rho ({\bf r}) g ({\bf k}, {\bf r})$. In the
absence of pion--source-medium interactions, so that $g ({\bf k},
{\bf r}) =1$, the integral is reduced to the Fourier transform of the
source distribution $\rho ({\bf r})$. In this limit we thus recover
the familiar result that the correlation function depends only on
${\bf q} \equiv {\bf k}_1 - {\bf k}_2$. In general, however, $f ({\bf
k}_1, {\bf k}_2)$ depends on the relative angle $\phi$ of the two
momenta ${\bf k}_1$ and ${\bf k}_2$; hence, the correlation function
also has this dependence.

In Sec. III, we present numerical results for a finite, complex
optical potential. However, a discussion of the black sphere limit
($\sigma_I \rightarrow -\infty$, $\rho (r) = \rho_o \theta (R-r) $)
serves as a useful illustration of the physics involved. Under these
circumstances the outgoing pion amplitude attenuates to zero when its
classical path transits the potential region. On the other hand, the
source distribution $\rho (r)$ is zero for $r > R$. We shall assume
that the pions are emitted in a thin spherical shell of width
$\delta$ outside $R$ and then compute the $\delta\rightarrow 0$
limit. The distortion factor $g({\bf k},{\bf r})$ is determined
purely by geometry:
\begin{equation}
g ({\bf k}, {\bf r}) = \theta \left({\hat {\bf k}} \cdot {\hat {\bf
r}}
+ \sqrt{1-{R^2 \over r^2}} \right)
\theta (r - R)\;.
\label{g}
\end{equation}
Equation~(\ref{f}) thus reduces to
\begin{equation}
f({\bf k}_1,{\bf k}_2)
={1 \over {2\pi}} \int d\Omega \ \theta ({\hat {\bf k}_1} \cdot {\hat
{\bf r}})
\ \theta ({\hat {\bf k}_2}
\cdot {\hat {\bf r}}) \ e^{i{\bf q} \cdot {\bf R}}\;.
\label{f1}
\end{equation}
The resulting correlation function can be evaluated numerically for
general $\phi$. It can also be calculated analytically in the
following two cases. For parallel pions, so that $\phi = 0$, $f$
becomes
\begin{equation}
f ({\bf k}_1, {\bf k}_2) = {1 \over i qR} \left[ e^{iqR} -1 \right]
\;,
\end{equation}
so that
\begin{equation}
C_2 ({\bf k}_1, {\bf k}_2) = 1 +
{4 \over q^2 R^2} \sin^2 \left({qR\over 2} \right) \;.
\label{c20}
\end{equation}
In this case, the correlation function is simply a function of the
dimensionless quantity $qR$, where $q \equiv \vert{\bf q}\vert$ is
the magnitude of the relative 3-momentum of the two pions. However,
for pions emitted back-to-back, so that $\phi = \pi$, $f$ vanishes.
Thus, $C_2 ({\bf k}_1, {\bf k}_2) \vert_{{\bf\hat{k}_2}=
-{\bf\hat{k}_1}} =1$ for all $q$. The attenuation of the emission
amplitude which penetrates the absorbing region destroys the
interference entirely.

The physical origin of the $\phi$ dependence can be readily
understood in this limit and is illustrated in Figs.~1 and 2. For
parallel pairs the deviation of the correlation function from unity
is caused by the constructive interference of the two amplitudes
corresponding to Figs.~1a and 1b. However, for antiparallel pairs,
the amplitude for the process in Fig.~2b, which would interfere with
the amplitude in Fig.~2a, is completely suppressed by absorption in
the source region; hence, in this case there is no enhancement.

We can now proceed to evaluate Eq.~(\ref{f1}) for arbitrary $\phi$
and total pion momentum $P \equiv \vert {\bf k}_1 + {\bf k}_2 \vert$.
For given $P$ and $\phi$, there is a minimum $q$ allowed
kinematically; that is,
\begin{equation}
q_{\rm min} = P\left({1-
\cos\;\phi \over 1+\cos\;\phi} \right)^{1/2} \;.
\label{qmin}
\end{equation}
As $\phi$ increases, $q_{\rm min}$ becomes non-negligible. At large
opening angles, $q_{\rm min}$ is large, and only a small correlation
between the pions remains --- regardless of any pion--source-medium
interactions. We therefore consider small opening angles, where
$q_{\rm min}$ is small, so that the suppression of the correlation
due to interactions with the source medium is easily detectable.

In Fig.~3, we show the correlation function for several $\phi$ at two
fixed values of $PR$ in the above black sphere limit. As shown in
Eq.~(\ref{c20}), the correlation function of parallel pairs goes to
two at zero $q$. For comparison, the solid line shows the correlation
function which results from ignoring final-state interactions. That
is, $g({\bf k}, {\bf r})$ is set to unity in Eq.~(\ref{f}), and a
uniform, spherical source of radius $R$ is assumed, so that
\begin{equation}
C_2^{\rm free} (qR) = 1+\left[ {3(\sin\; qR - qR \;\cos\; qR) \over
(qR)^3}
\right]^2 \;.
\end{equation}
The two correlation functions are independent of $\phi$ and $P$, yet
the scales with which they fall off in $qR$ are different. In the
{\it black} sphere case the correlation of parallel pairs at
intermediate $q$ is {\it enhanced}. This result can be understood
physically as follows: when the source region is black, only pion
pairs emitted from the front surface --- as shown in Fig.~1 --- are
detected; the pairs emitted from opposite sides as illustrated in
Fig.~4 are suppressed. The effective source size is thus {\it
smaller} than it would be if all the pairs were sampled, leading to
an enhanced correlation at intermediate $q$. When the opening angle
between the pions becomes non-zero, the correlation function is
suppressed relative to the parallel pion case, due to the final-state
interactions. For $PR=10\, (20)$, corresponding to $P \approx 400 \,
(800)$~MeV for a source radius of $R\approx5$~fm, Fig.~3 shows the
suppression of the correlation function as $\phi$ increases from
$0^o$ (dashed line) to $5^o$ (crosses), $10^o$ (squares), and $20^o$
(pluses).

Based on the above discussion, we hypothesize that these absorption
effects should be quite general and thus present in fermion
interferometry as well.

\section{Numerical results and phenomenological consequences}
\label{secnumer}

In this section we compute the correlation function ratio,
Eq.~(\ref{Cstart}), for the finite, complex optical potential of
Eq.~(\ref{potent}). For illustrative purposes, we choose simple forms
for the source density $\rho(r)$, a Gaussian $\rho (r) = \rho_o \exp
(-r^2 / {R_G}^2)$ and a uniform sphere $\rho (r) = \rho_o \theta (R_S
- r) $, with $R_S = \sqrt{2.5} R_G$. The density parameters are
chosen so that the two source functions have the same rms radii and
central densities. Note that the integral $\int d^3 r \rho (r)$
differs in the two densities chosen. We equate the central densities
of the two sources, rather than the number of particles $\int d^3 r
\rho (r)$ they contain, as we believe the central density better
characterizes the heavy ion collision. We presume that the fireball
resulting from the heavy-ion collision is baryon-rich for simplicity,
as is the case for heavy ion experiments in the nuclear stopping
regime, such as those at the AGS \cite{exp,exp1}. We thus choose
$\rho_o =0.17~{\rm fm}^{-3}$.

Given $\rho(r)$, we need to estimate the constants $\sigma_R$ and
$\sigma_I$. In our treatment of $U(x)$, $\sigma_R$ and $\sigma_I$ are
proportional to the effective, forward $\pi N$ scattering amplitude
in the fireball. The $\pi N$ amplitude in free space is well-known
\cite{particle_data}: due to 3-3 resonance formation, the imaginary
part of the amplitude has a large peak near a pion laboratory energy
of 200~MeV, around which the real part changes the sign. Below and
above this energy region, the amplitude varies less and is of smaller
magnitude. It involves many $\pi N$ resonances at higher energies.

The construction of an effective $\pi N$ amplitude in nuclei from the
free-space amplitude is an involved procedure, but the issues are
well understood \cite{eis_kol}. The main issues involved in its
construction are as follows. 1) $\Delta$ resonance formation depends
on the {\it local} relative momentum of the $\pi-N$ system, so that
in the medium a minimally non-local optical potential is required. 2)
Nuclear pion absorption involves more than one nucleon, so that the
nucleon correlations in the medium become important. 3) Higher-order
scattering contributions are now possible as well, and these also
involve nucleon correlations. In view of the uncertainties in
averaging over the initial source states and summing over the final
states in constructing the potential in the fireball, we assume that
these complications can be effectively included by varying the
parameter values in our local potential.

Since we use a simple model to illustrate the essential physics that
emerge, we first choose a representative parameter set for $\sigma_R$
and $\sigma_I$ and vary it over a wide range. The representative set
is based on 1) the free-space $\pi N$ scattering amplitude, taking
into account 2) nuclear pion absorption and 3) the non-observation of
the initial and final states of the pion source, which acts to
increase the pion's effective mean-free path \cite{glauber,kys}. We
assume that the form of Eqs.~(11) and (12) is preserved under the
inclusion of this last effect, yet this need not be so. Finally, the
representative set we choose is $(\sigma_R,\sigma_I) = (\pm 2,
-2)~{\rm fm}^2$, which corresponds to a characteristic absorption
length of $(\sigma \rho_o)^{-1} = 1.5$~fm in nuclear matter. In order
to explore a wider range of parameter values, we multiply this
reference value by a scale factor $c$ such that $(\sigma_R, \sigma_I)
= (\pm 2, -2)c~{\rm fm}^2$ and vary $c$. To ensure that our
examination is not prejudiced by our choice of the imaginary to real
ratio, we also examine the cases $\sigma_R =0$ and $\sigma_I =0$.

In general, the correlation function depends on the kinematics of the
pion pair --- ${\bf k}_1, {\bf k}_2$ --- as well as on the parameters
characterizing the source --- $R, \sigma_R, \sigma_I$. For a
spherically symmetric source, if final-state interactions are
ignored, the correlation function depends only on $q$ and $R$.
However, if $U \neq 0$, it depends on the opening angle $\phi$ and
the total pair momentum $P$ as well. In this section, we discuss
these new dependences in turn.

As discussed previously, we expect the correlation function to be
suppressed at large opening angles as a consequence of including
absorption in the source region. Our calculations with a finite
optical potential show that to be indeed the case. In Fig.~5 we plot
the opening angle dependence of the correlation function at
$P=400$~MeV and $q=40$~MeV for both a uniform source of $R_S=10$~fm
and a Gaussian source of $R_G=5$~fm. The $c=0$ results --- here there
are no final-state interactions --- do not depend on $\phi$, yet
there is always a significant suppression of $C_2$ as $\phi$
increases for finite $\sigma_I$, in qualitative agreement with the
black sphere results. We observe the same behavior if we fix
$\sigma_I$ and vary $\sigma_R$, or vice versa, rather than varying
them together. To wit, {\it absorption by the source leads to
suppression of the correlation at finite opening angles}.

If one averages the correlation function over the opening angle, the
results turn out to be rather insensitive to $P$. We therefore fix
$P$ and investigate the sensitivity of the opening-angle-averaged
$C_2$ to the optical potential. The range of $\phi$ is restricted
kinematically:
\begin{equation}
\phi < {\rm arccos} \left[ {P^2 - q^2 \over P^2 +q^2} \right]\;.
\end{equation}
Thus, for fixed $P$, as $q$ increases, one includes an ever larger
proportion of large opening angle pairs in $C_2$, the correlation of
which are suppressed. Angle-averaging thus results in a rather
reduced correlation in the intermediate and large $q$ range. At the
lowest $q$, only parallel pairs are included, and the correlation
approaches the Bose-Einstein limit, $C_2 = 2$, as $q\rightarrow 0$,
producing a non-Gaussian shape in the low $q$ region. If one fits a
standard Gaussian form to the correlation function,
\begin{equation}
C_2 = 1+\lambda \;{\rm exp} \left[ -q^2{R_e}^2 \right]\;,
\label{stgaus}
\end{equation}
as is done with experimental data \cite{exp}, to extract the
effective source size $R_e$ and the ``coherence parameter''
$\lambda$, one finds a smaller effective source size $R_e < R$ and
$\lambda < 1$. The absorption of pions in the source region therefore
leads to a small effective $R_e$ and $\lambda$, as well as to a
non-Gaussian shape of the correlation function in the low $q$ region.

In Fig.~6 we show the opening-angle-averaged correlation functions
for both sources with several values of $c$. Note that $P=400$~MeV
and $R=10(5)$~fm for the uniform (Gaussian) source. For each source,
we show the results for both signs of $\sigma_R$. Three curves are
shown in each panel corresponding to $c=0$ (dashed lines), 1 (solid
lines), and 10 (dot-dash lines). These results indicate a small but
significant change in the shape of the correlation functions once the
final-state interactions are turned on. Indeed, the medium-modified
correlation function displays a non-Gaussian shape, especially at low
$q$, even for a Gaussian source.

Such non-Gaussian correlation functions have been seen
\cite{exp,exp1,E665}. For comparison, we show the E802 data for the
$\pi^-$ correlations in $14.6A~{\rm GeV/c}$ ${}^{28}{\rm Si} +
{}^{197}{\rm Au}$ collisions \cite{exp1} together with our
calculation in Fig.~7. The data can be fit to the form in
Eq.~(\ref{stgaus}) with $R_e = 3.42 \pm 0.26$~fm and $\lambda = 0.65
\pm 0.07$, yet the static Gaussian model we use also reproduces the
correlation function if one assumes $R_G \approx 5$~fm and $c \neq
0$. The solid line in the figure is calculated assuming $P = 400$~MeV
and $c=10$. The results for $P=600$~MeV are almost identical. The
agreement with data can be improved if we vary both the optical
potential and $R_G$. However, our goal here is not a quantitative fit
to the data --- especially in light of our simple assumptions
regarding the static source and its interaction --- but, rather, a
qualitative insight into how non-Gaussian shapes can be generated.
Note that Fig.~7 also suggests that a study of the opening-angle
dependence may be a better way to infer the extent of
pion--source-medium interactions.

If $\vert\sigma_I\vert$ is small, one may be able to detect the
correlation effects induced by the real part of the optical
potential. The extra phase introduced by the final-state interaction,
$\exp (-i \sigma_R t)$, modifies each of the pion wavefunctions and
leads to an interference pattern in the correlation function. Such
effects can be seen most clearly if a sharp edge is present in the
source and the damping factor $\exp (-\vert\sigma_I\vert t)$ is
small. In Fig.~8, we show the $\phi$-averaged correlation functions
for a uniform spherical source with several values of $\sigma_R$ and
two values of $\sigma_I$. A large oscillation is observed in Fig.~8a
when $\sigma_I = 0$ and $\sigma_R$ is of roughly $-1.5~{\rm fm}^2$.
The correlation function becomes as large as 4.5 at $q=100$~MeV!
Note, however, that we have not excluded bound-state forming
parameters for $\sigma_R < 0$ and small $\sigma_I$. The oscillations
damp out quickly as $\sigma_I$ grows large, as shown in Fig.~8b.

At fixed $P$, $q$, and $R$, one can compare the values of $C_2$ for
different $\sigma_R$ and $\sigma_I$. Fig.~9a shows the variations of
$C_2$ with respect to $\sigma_R$, for various fixed $\sigma_I$. In
the upper panel, $\sigma_I=0$, whereas $\sigma_I =-2$ and $-4~{\rm
fm}^2$ --- the solid and dashed lines, respectively --- in the lower
panel. The interference pattern is clearly seen at $\sigma_I = 0$,
yet only a small variation is observed for a more realistic value of
$\sigma_I$. Note, however, the sudden jump in $C_2$ when $\sigma_R$
changes sign. Similarly, we study $C_2$ as a function of $\sigma_I$
for various $\sigma_R$ in Fig.~9b. That is, we choose $\sigma_R=\pm
2~{\rm fm}^2$ --- the dashed and solid lines --- and 0 --- the
dot-dashed line. The interference induced by $\sigma_R$ enhances or
suppresses the correlation for $\sigma_R < 0$ or $\sigma_R > 0$ if
$\vert\sigma_I \vert$ is less than about 2~fm$^2$.

Even though our model is currently too simple to take its comparison
with experiment seriously, the generation of modifications to the
correlation function's shape by the pion--source-medium interaction
is significant and robust. From our model the following simple
picture emerges: the inclusion of pion--source-medium interactions
leads to 1) a suppression of the correlation as the opening angle
increases, 2) an angle-averaged correlation function which can
deviate from a Gaussian shape at low $q$, as well as 3) possible
additional interference effects from the real part of the potential.
The last effect is suppressed by the inclusion of absorption, and
when $\sigma_I$ exceeds or is of the order of the value in normal
nuclear matter the correlation becomes insensitive to the real part.
If one fits the correlation data with a simple Gaussian form,
Eq.~(\ref{stgaus}), these effects will lead to a small {\it apparent}
source radius and a small coherence parameter, $\lambda < 1$.

\section{Conclusions}
\label{secconcl}

Using a simple model for the pion--source-medium interaction, in the
eikonal approximation, we have demonstrated that an interesting
momentum dependence in the two-pion momentum correlation function is
generated. Specifically, when absorptive interactions are finite, the
pion correlation function decreases rapidly as the opening angle
between the pions increases. Our model is a simple one, yet this
behavior should not be sensitive to the assumptions we have made.
Pion-interferometry analyses have usually been performed as a
function of only two of the two-pion momentum variables, implicitly
assuming that the correlation function is independent of the opening
angle. Our discovery of a strong opening-angle dependence thus
suggests that analyses should be carried out explicitly in terms of
the opening angle as well. Note, moreover, that the
opening-angle-averaged pion correlation function exhibits a
non-Gaussian shape when pion--source-medium interactions are
included, even for a Gaussian source. This provides an alternative
explanation for the coherence factor $\lambda$, which is usually
invoked to fit experimental data.

\acknowledgments

We thank S.~E.~Koonin and M.~Gyulassy for stimulating discussions,
and M.~Macfarlane and B.~Serot for comments on the manuscript. This
research is supported in part by the U.~S.~National Science
Foundation, Grants No. PHY90-13248 and PHY91-15574, at Caltech, by
the U.S. Department of Energy, Grant No. DE-FG02-87ER40365 at Indiana
University and Grant No. DE-FG03-87ER40347 at California State
University, Northridge, and by the Grant-in-Aid for Scientific
Research, Japan Ministry of Education and Culture, No. C06640394.

\begin{figure}
\caption{Emission of a parallel pair of pions from the source region.
The amplitude in a) interferes constructively with the exchange
amplitude in b).}
\label{fig1}
\end{figure}

\begin{figure}
\caption{Emission of an anti-parallel pair of pions from the source
region. The amplitude in a) would interfere constructively with the
exchange amplitude in b), but the latter is suppressed by absorption
in the source region.}
\label{fig2}
\end{figure}

\begin{figure}
\caption{Two-particle correlation from a black sphere source of
radius $R$ as a function of the relative momentum $q = \vert{\bf k}_1
- {\bf k}_2 \vert$. The total momentum $P = \vert{\bf k}_1 + {\bf
k}_2 \vert$ is $10/R$ in a) and $20/R$ in b). The solid lines
indicate the results with no pion--source-medium interactions. The
source region is assumed to be spherical and uniform. When the source
medium is absorptive, the correlation function also depends on the
opening angle $\phi$ between the particles. Shown are the results for
$\phi = 0^o$ (dashed lines), $5^o$ (crosses), $10^o$ (squares), and
$20^o$ (pluses). For fixed $P$ and $\phi$, there is a minimum $q$,
Eq.~(\protect{\ref{qmin}}), allowed kinematically.}
\label{fig3}
\end{figure}

\begin{figure}
\caption{Emission of a parallel pair of particles from opposite sides
of the source region. This process is suppressed compared to the
emission in Fig.~1 if the source region is absorptive.}
\label{fig4}
\end{figure}

\begin{figure}
\caption{Two-pion correlation as a function of the opening angle
$\phi$ between the particles. The relative momentum is fixed at $q =
40$~MeV, and the total momentum at $P = 400$~MeV. The optical
potential is chosen with $(\sigma_R, \sigma_I) = (\pm 2, -2) c~{\rm
fm}^2$, and the results are shown here for a range of $c$. If the
final-state interaction is ignored, so that $c=0$, $C_2$ does not
depend on $\phi$, as indicated by the solid lines. When absorption is
present, the correlation decreases as $\phi$ increases. Note that
$c=1,\, 10$ correspond to the dashed and the dot-dashed lines,
respectively. Results for both an uniform sphere of radius $R_S
=10$~fm, in a) and c), and a Gaussian source of $R_G =5$~fm, in b)
and d), are shown with both signs of $\sigma_R$.}
\label{fig5}
\end{figure}

\begin{figure}
\caption{Two-pion correlation function averaged over the pions'
opening angle, as a function of $q$. The total pion momentum is
$P=400$~MeV. Results for three different final-state interactions are
shown, and, as in Fig.~5, $c=0$, 1, and 10 correspond to the dashed,
solid, and dot-dashed lines, respectively.}
\label{fig6}
\end{figure}

\begin{figure}
\caption{Comparison of the pion correlation function calculated using
a static Gaussian source model ($R_G= 5$~fm) and the E802 data,
corrected for acceptance and Coulomb effects, for the $\pi^-$
correlation in $14.6A$~GeV/c ${}^{28}{\rm Si} + {}^{197}{\rm Au}$
collisions \protect{\cite{exp1}} with $q$. The solid line indicates
the results using $\sigma = -20 -20i~{\rm fm}^2$ ($c=10$), whereas
the dashed line represents the non-interacting ($c=0$) results. In
our calculation $P = 400$~MeV, but the results are insensitive to
$P$.}
\label{fig7}
\end{figure}

\begin{figure}
\caption{Opening-angle-averaged pion correlation function as a
function of $q$ for various $\sigma_R$ and a uniform spherical source
of $R_S =10$~fm. $P$ is fixed at 400~MeV, and in a) $\sigma_I =0$,
whereas in b) $\sigma_I=-2~{\rm fm}^2$. The curves correspond to
different values of $\sigma_R$; that is, the results with
$\sigma_R=-3$, $-1.5$, 0, and 3~fm$^2$ are denoted by dashed,
dot-dashed, solid, and dotted lines, respectively. A large
oscillation is seen for $\sigma_R \sim -1.5~{\rm fm}^2$ if $\sigma_I
= 0$, but very little sensitivity to $\sigma_R$ is observed for more
realistic $\sigma_I$.}
\label{fig8}
\end{figure}

\begin{figure}
\caption{Dependence of the opening-angle-averaged pion correlation
function on a) $\sigma_R$ and b) $\sigma_I$. The calculation is
performed with $P=400$~MeV and $q=40$~MeV for a uniform spherical
source of $R_S =10$~fm. In a), the upper panel shows the results for
$\sigma_I =0$, whereas the lower panel shows those for $\sigma_I=-2$
and $-4~{\rm fm}^2$ --- the solid and dashed lines, respectively. In
b), the results for $\sigma_R = -2$, 0, 2~fm$^2$ are denoted by
solid, dot-dashed, and dashed lines. The oscillations seen in Fig.~8
give rise to a large modification of the correlation at small
$\vert\sigma_I \vert$.}
\label{fig9}
\end{figure}

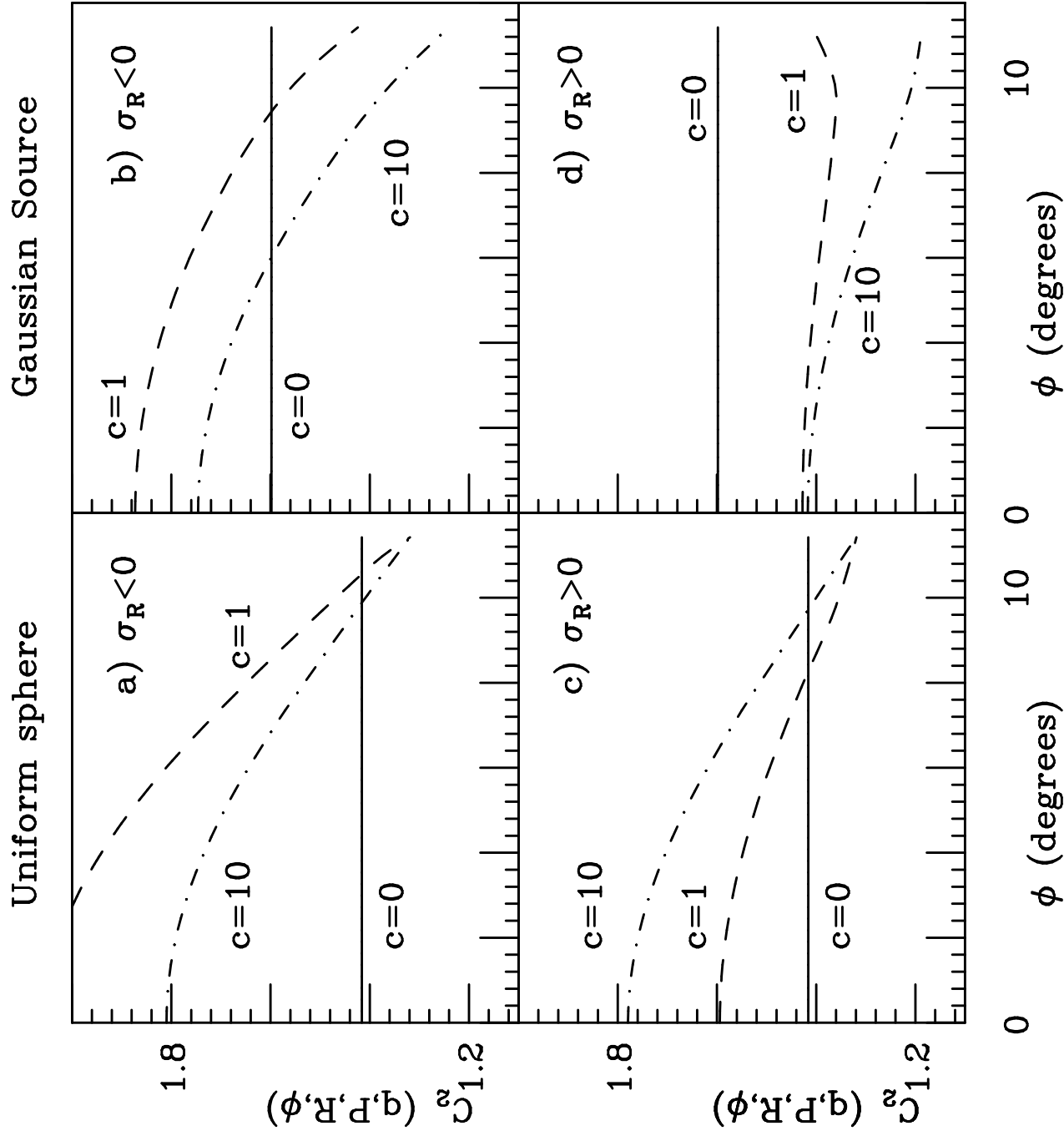
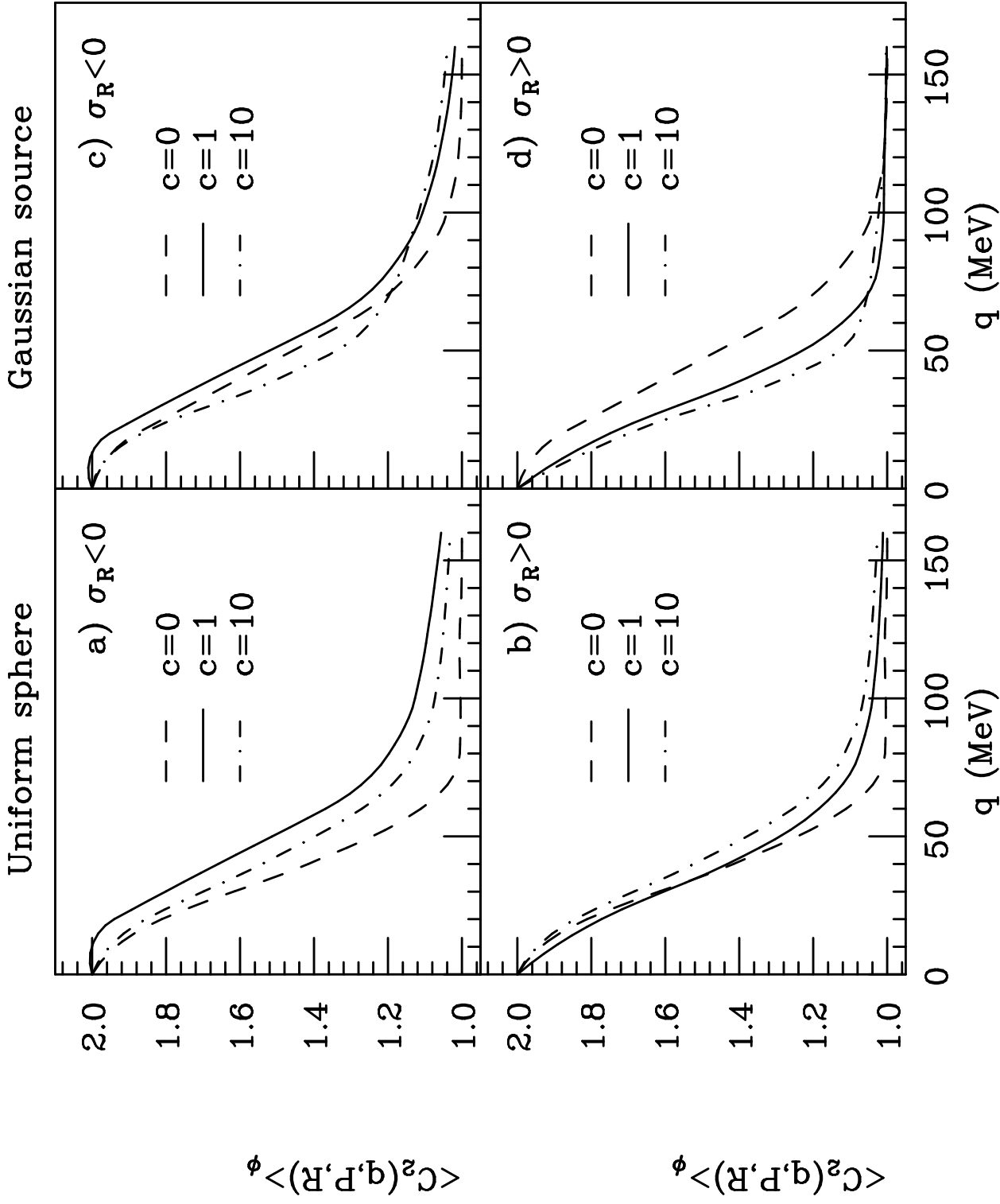
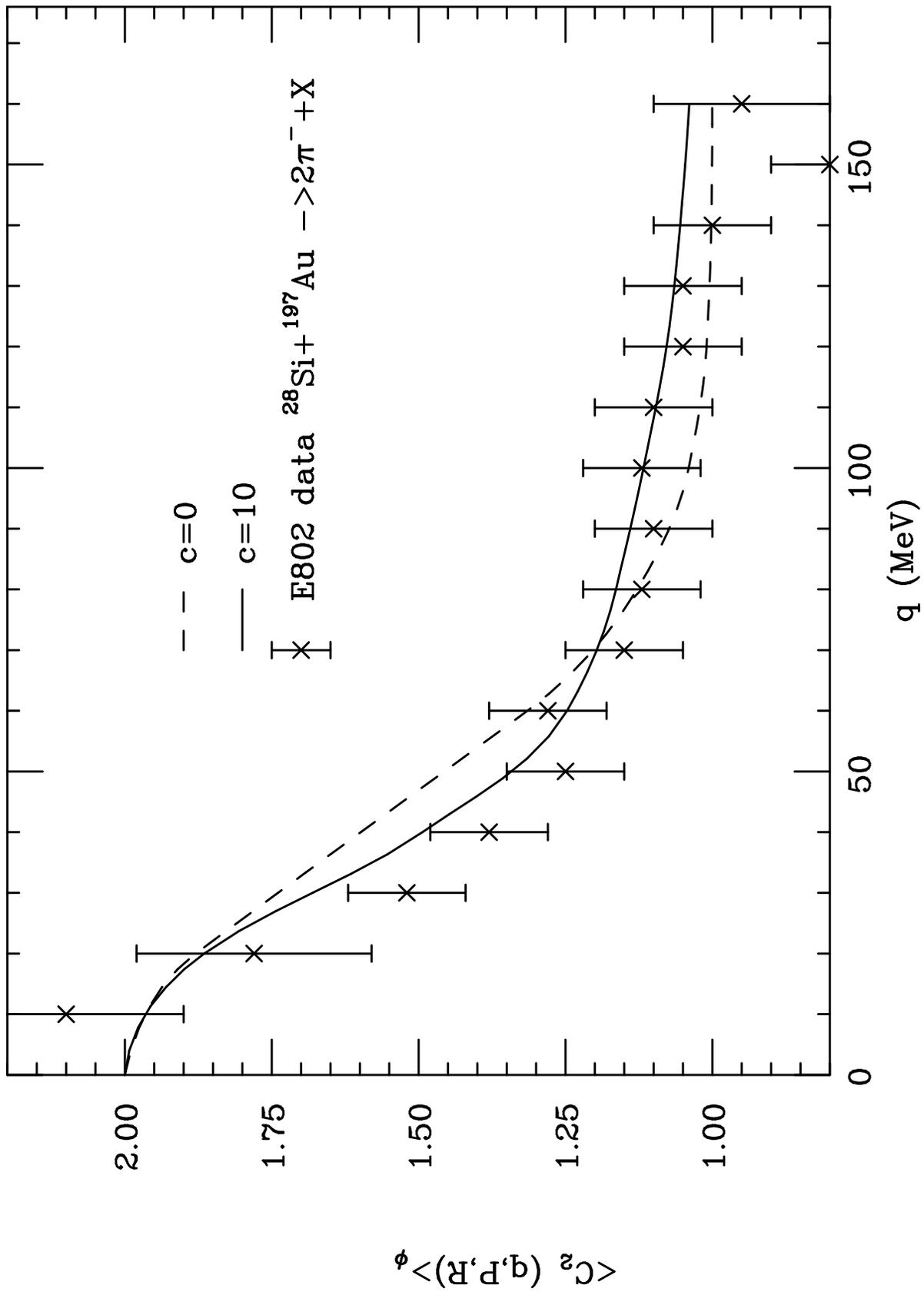
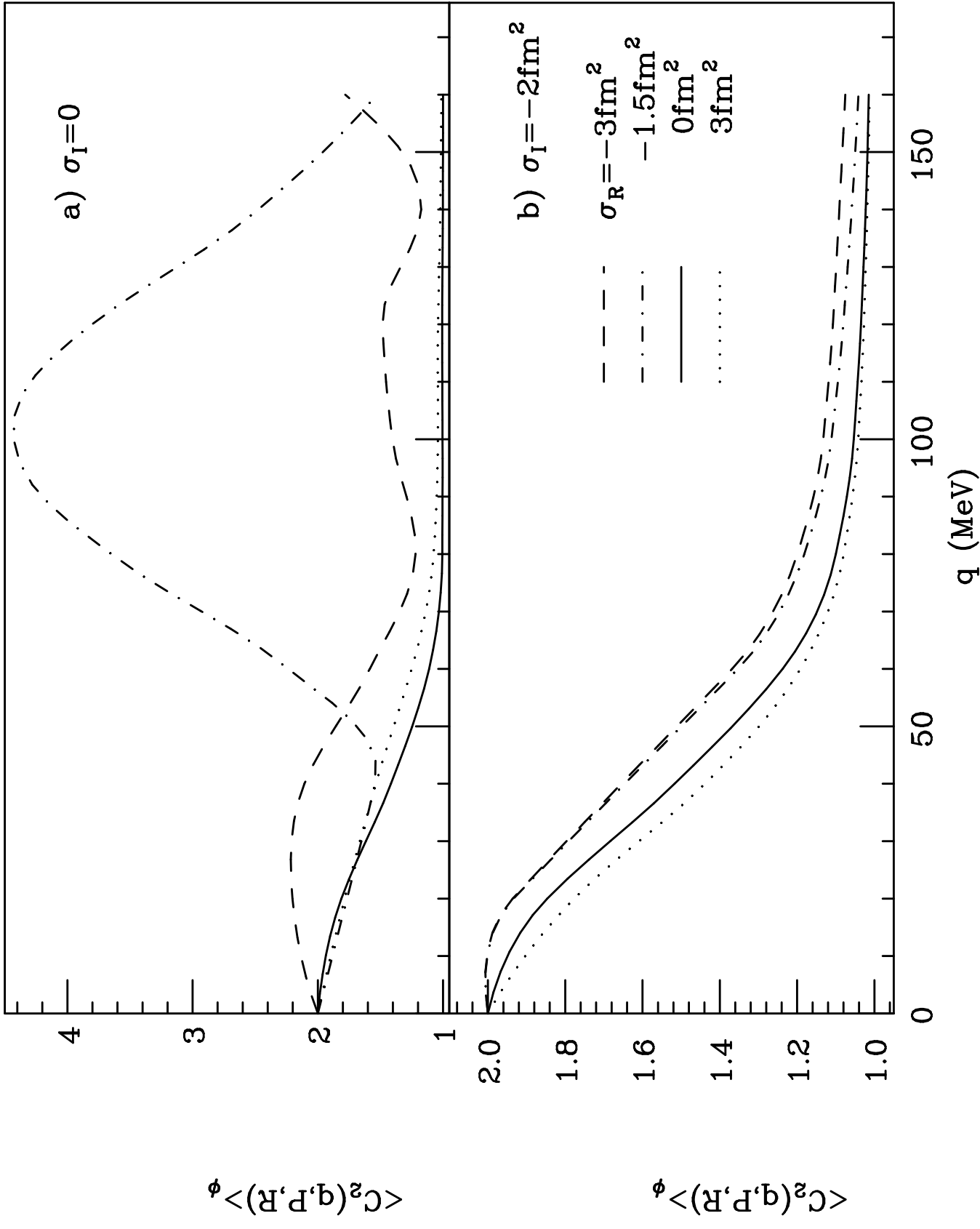
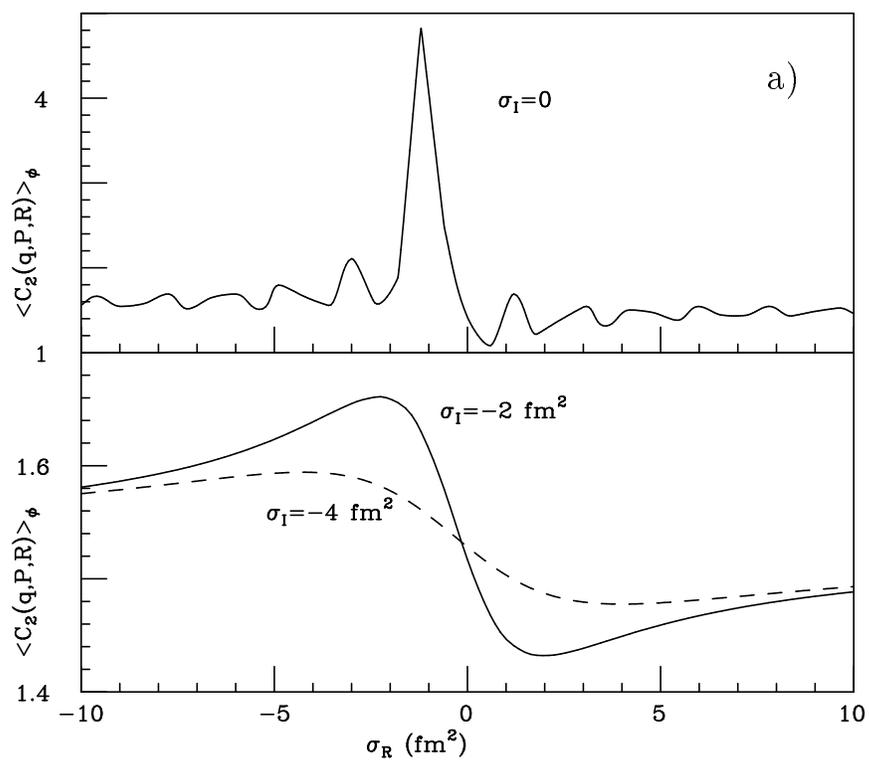
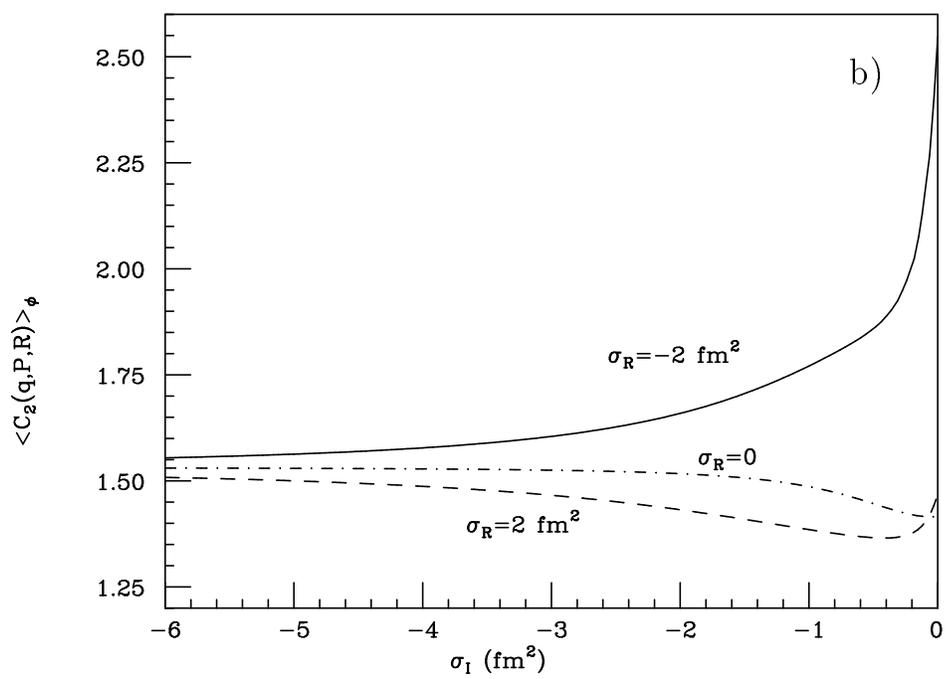
\end{document}